\documentclass{ragtime} 
\title[Chaotic motion in the Johannsen-Psaltis spacetime]%
      {Chaotic motion in the Johannsen-Psaltis\\ spacetime}

\author[O. Zelenka and 
        G. Lukes-Gerakopoulos    
     						]
       {Ond\v{r}ej Zelenka\at[]{1,2,a} 
        and Georgios Lukes-Gerakopoulos\at[]{1,b} 
        \\%
        \ins{1}Astronomical Institute of the Academy of Sciences of the Czech Republic,\splitins[1]
        Bo\v{c}n\'{i} II 1401/1a, CZ-141 31 Prague, Czech Republic\\
        \ins{2} Institute of Theoretical Physics, Faculty of Mathematics and Physics,\splitins[2]
Charles University, CZ-180 00 Prague, Czech Republic\\
        \ins{a}\Email{ondrzel@gmail.com} 
        \ins{b}\Email{gglukes@gmail.com}} 

\providecommand{\defeq}{:=} 
\providecommand{\pder}[2]{\frac{\partial #1}{\partial #2}} 
\providecommand{\N}{\mathbb{N}} 

\begin{document}

\begin{abstract}
The Johannsen-Psaltis spacetime is a perturbation of the Kerr spacetime designed
to avoid pathologies like naked singularities and closed timelike curves. This
spacetime depends not only on the mass and the spin of the compact object, but
also on extra parameters, making the spacetime deviate from Kerr; in this work
we consider only the lowest order physically meaningful extra parameter. We use
numerical examples to show that geodesic motion in this spacetime can exhibit
chaotic behavior. We study the corresponding phase space by using Poincar\'{e}
sections and rotation numbers to show chaotic behavior, and we use Lyapunov
exponents to directly estimate the sensitivity to initial conditions for chaotic
orbits.
\end{abstract}


\section{Introduction}\label{intro}

We study the geodesic motion in a family of spacetimes constructed by 
\cite{JohannsenPsaltis11}. The corresponding metric is characterized by an
infinite number of parameters, i.e. the mass $M$, the spin $a$ and a series of
deviation parameters $\epsilon_k$, where $k\in\N_0$. However, in this work we
constrain ourselves to the lowest order of the unconstrained parameters, which
is $\epsilon_3$.

The Johannsen-Psaltis (JP) metric was designed to be a perturbation of the Kerr
spacetime, which is of great astrophysical interest. The so-called
\textit{no-hair theorem} \citep[see, e.g.,][]{Carter71} states that the class of
uncharged black-hole exterior solutions which are axisymmetric and don't violate
causality (i.e. no closed timelike curves) consists of a discrete set of
continuous families, each depending on at least one and at most two independent
parameters. No other externally observable parameters are required for this
description. Typically, the Kerr spacetime is assumed to describe a black hole
\citep{Rico13}. Kerr black holes are parametrized by their mass $M$ and their
angular momentum $a$. However, there is yet to be a proof if black holes are
indeed described by the Kerr paradigm. Therefore, it would be of great
astrophysical interest to test this conjecture by observing black hole
candidates through electromagnetic and gravitational wave signals.

The Kerr spacetime is axisymmetric and stationary, but one special feature of this
spacetime is that it has an extra "hidden" symmetry that makes geodesic motion
in such a background correspond to an integrable system \citep{Carter68}. There
are spacetimes that deviate from Kerr by a deformation parameter, these
spacetimes are called in the bibliography non-Kerr spacetimes
\citep[see, e.g.,][]{Bambi17}. These non-Kerr spacetimes do not usually possess
the symmetry that the Kerr spacetime does, making geodesic motion correspond to
a non-integrable system. As a result, geodesic motion in such spacetimes
exhibits chaotic behavior, which is the topic of our study.

The organization of the article is as follows: in section \ref{sec:theory} we 
describe the basics of geodesic motion, deterministic chaos in dynamical systems
and some of the properties of the JP spacetime. In section \ref{sec:numerics} we
use numerical examples to show that the JP metric doesn't correspond to an
integrable system. Section \ref{sec:conclusion} summarizes our main findings.
Note that geometric units are employed throughout the article, ${G=c=1}$. Greek
letters denote the indices corresponding to spacetime and the metric signature
is $(-,+,+,+)$.

\section{Geodesic motion and chaos}\label{sec:theory}

The line element of a rapidly spinning black hole introduced in 
\cite{JohannsenPsaltis11} reads in Boyer-Lindquist-like coordinates 
\begin{equation}     \label{eq:metric}
\mathrm{d}s^2 = g_{tt}\mathrm{d}t^2 + g_{rr}\mathrm{d}r^2 +
g_{\theta\theta}\mathrm{d}\theta^2 + g_{\phi\phi}\mathrm{d}\phi^2 + 2g_{t\phi}\mathrm{d}t\mathrm{d}\phi \quad,
\end{equation}
where the metric components $g_{\mu\nu}$ \citep{JohannsenPsaltis11} are
\begin{subequations}
\begin{alignat}{3}
&g_{tt} &&= -&&\left(1+h\right)\left(1-\frac{2Mr}{\Sigma}\right) \quad,\\
&g_{t\phi} &&= -&&\frac{2aMr\sin^2\theta}{\Sigma}\left(1+h\right) \quad,\\
&g_{\phi\phi} &&= &&\frac{\Lambda\sin^2\theta}{\Sigma} + ha^2\left(1 + \frac{2Mr}{\Sigma}\right)\sin^4\theta \quad,\\
&g_{rr} &&= &&\frac{\Sigma\left(1+h\right)}{\Delta + a^2h\sin^2\theta} \quad,\\
&g_{\theta\theta} &&= &&\Sigma \quad,
\end{alignat}
\end{subequations}
and the metric functions are
\begin{subequations}
\begin{alignat}{2}
&\Sigma &&= r^2 + a^2\cos^2\theta \quad,\\
&h &&= \sum_{k=0}^{\infty}\left(\epsilon_{2k}+\epsilon_{2k+1}\frac{Mr}{\Sigma}\right)\left(\frac{M^2}{\Sigma}\right)^k \quad, \label{eq:h}\\
&\Delta &&= r^2 + a^2 - 2Mr \quad,\\
&\omega^2 &&= r^2 + a^2 \quad,\\
&\Lambda &&= \omega^4 - a^2\Delta\sin^2\theta \quad.
\end{alignat}
\end{subequations}

The function $h\left(r,\theta\right)$ is what causes the deviation from the Kerr
metric. Namely, setting $\epsilon_k=0\quad\forall k\in\N_0$ gives the Kerr metric.
The parameters $\left(\epsilon_k\right)_{k=0}^{\infty}$ are, however, constrained.
As explained in detail in \citep{JohannsenPsaltis11}, we have to set
$\epsilon_0$ = $\epsilon_1$ = 0 and the parameter $\epsilon_2$ is constrained by
observational constraints on weak-field deviations from general relativity
\citep{JohannsenPsaltis11}, i.e. $\left|\epsilon_2\right| \leq 4.6\cdot 10^{-4}$.
We therefore set $\epsilon_2=0$ as well and limit ourselves to the lowest order
remaining parameter, which is $\epsilon_3$, and set all the higher order
parameters $\epsilon_k = 0 \quad\forall\quad k \geq 4$.

The proper time $\tau$ defined as $\mathrm{d}\tau^2 = -g_{\mu\nu}\mathrm{d}x^\mu\mathrm{d}x^\nu$
is employed as the evolution parameter. The geodesic motion of a free particle
of rest mass $m$ is then generated by the Lagrangian \citep[see, e.g.,][]{Rindler06}
\begin{equation}\label{eq:L}
\mathcal{L}\left(x^\mu,\dot{x}^\mu\right) = \frac{m}{2}g_{\mu\nu}\dot{x}^\mu\dot{x}^\nu \quad,
\end{equation}
where dot denotes a derivative with respect to the proper time. Due to the
preservation of the four-velocity $g_{\mu\nu}\dot{x}^\mu\dot{x}^\nu = -1$ along
a geodesic orbit $\mathcal{L} = -m/2$ is a constant. The corresponding canonical
momenta are
\begin{equation}
p_\mu = \pder{\mathcal{L}}{\dot{x}^\mu} = mg_{\mu\nu}\dot{x}^\nu
\end{equation}
and performing the Legendre transform gives the Hamiltonian
\begin{equation}
\mathcal{H} = \frac{1}{2m}g^{\mu\nu}p_\mu p_\nu \quad.
\end{equation}
The JP metric functions are independent of the parameters $t$ and $\phi$, i.e.
it is stationary and axisymmetric, therefore the energy $E \defeq -p_t$ and 
and the component of the angular momentum $L_z \defeq p_\phi$ are integrals of
motion. This allows us to restrict our study to the meridian plane generated by
the polar-like coordinates $\left(r,\theta\right)$ and move to a simpler system
of two degrees of freedom. Namely, one has to merely replace
\begin{equation}
\dot{t} = \frac{-g^{tt}E+g^{t\phi}L_z}{m},\quad \dot{\phi} = \frac{-g^{t\phi}E+g^{\phi\phi}L_z}{m}
\end{equation}
in the equations of motion to reduce the system. The motion in the resulting
reduced system is characterized by the Newtonian-like two-dimensional effective
potential
\begin{equation}
\left(p_r\right)^2 + \frac{g_{rr}}{g_{\theta\theta}}\left(p_\theta\right)^2 =
-V_\textrm{eff} \defeq -g_{rr}\left(1+\frac{g_{\phi\phi}E^2 + g_{tt}L_z^2 + 2g_{t\phi}EL_z}{g_{tt}g_{\phi\phi}-g_{t\phi}^2}\right) \quad.
\end{equation}
For $p_\theta = p_r = 0$ the roots of this effective potential $V_\textrm{eff}=0$
form a curve in the meridian plane, which is called the \textit{curve of zero velocity} (CZV).


In the Kerr case, an extra "hidden symmetry" exists\footnote{For more details on
this symmetry see \citep{{Markakis14}} and references therein.}, giving rise to 
the Carter constant $\mathcal{K}$ \citep{Carter68}. This constant, along with
$E$, $L_z$ and $\mathcal{H}$, are independent and in involution, therefore
geodesic motion in the Kerr spacetime background corresponds to an integrable
system and trajectories of the reduced system lie on a family of two-dimensional
\textit{invariant tori}. These orbits oscillate in both degrees of freedom with
their respective characteristic frequencies $\omega^r$ and $\omega^\theta$; their
ratio $\omega = \omega^r/\omega^\theta$ is called the \textit{rotation number}
and it is useful for the classification of orbits. If $\omega$ is rational, the
torus is called \textit{resonant} and it hosts an infinite number of periodic
orbits. If $\omega$ is irrational, the motion is called \textit{quasiperiodic}
and each orbit on the torus covers it densely.

When a perturbation is applied to such an integrable system, all the resonant
tori are destroyed. According to the \textit{KAM theorem} \citep{Meiss92},
however, most of the non-resonant tori survive in the perturbed system for
small perturbations; these are called KAM tori. According to the
\textit{Poincar\'{e}-Birkhoff theorem} \citep{LichtenbergLieberman92}, where there
was a resonant torus, an even number of periodic trajectories survives in the
perturbed system, half of them stable and half unstable. We use a Poincar\'{e}
surface of section to display the phase space structure of the system. We define
a surface in the phase space and plot the intersections of the orbits with the
surface. Invariant tori correspond to circles in the surface of section. These
form the \textit{main island of stability} around a stable fixed point in the
center.

Near the now destroyed resonant tori, quite a different structure arises. Around
the stable periodic points (corresponding to surviving stable periodic orbits),
smaller islands of stability arise, forming together with the unstable
points (corresponding to surviving unstable periodic orbits) \textit{Birkhoff chains}. 
These unstable periodic points lie between the aforementioned islands of stability.
From the unstable points emanate asymptotic manifolds, there are stable and
unstable branches. The branches of the same type cannot cross each other, which
results in very complicated structures in the phase space. These complicated
structures are the driving engines of deterministic chaos.

\begin{figure} [t]
\begin{center}
\graphicspath{{}}
\input{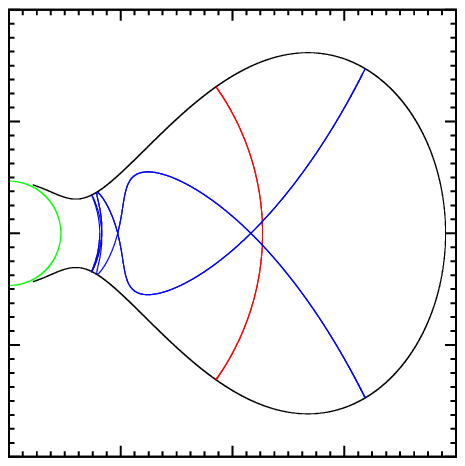}
\graphicspath{{}}
\input{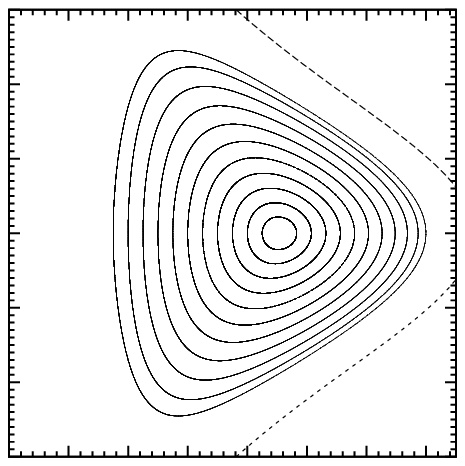}
\end{center}
\caption{\label{fig:projection_surface}
Left panel: projections of orbits on the meridian plane - the period-1 orbit (red),
the 5/7 periodic orbit (blue), the event horizon (green), the CZV (black). 
Right panel: The main island of stability on the Poincar\'{e} section
$\theta=\frac{\pi}{2}$.}
\end{figure}

An effective tool to analyze types of motion on a Poincar\'{e} section of a
non-integrable system of two degrees of freedom is the angular moment
$\nu_\vartheta$, known in the literature as the rotation number
\citep[see, e.g,][]{Voglis98,Voglis99}. We denote the central fixed point of
the main island of stability $u_c$ and the $n$-th crossing of the surface of
section by the orbit $u_n$. We define rotation angles
\begin{equation}
\vartheta_n \defeq \mathrm{ang}\left[u_{n+1}-u_c,u_n-u_c\right]
\end{equation}
and the angular moment as
\begin{equation}
\nu_\vartheta = \lim_{N\to\infty} \frac{1}{2\pi N}\sum_{n=1}^N\vartheta_n \quad.
\end{equation}
The dependence of this angular moment on the distance of the initial condition
from the central fixed point is called the \textit{rotation curve}. 
In an integrable system, such as the Kerr spacetime, the rotation curve is 
strictly monotonous, but in a non-integrable system, it has non-monotonic
variations when passing through chaotic zones, and plateaus when passing through
islands of stability.

In order to quantify sensitivity to initial conditions, which is a property of 
chaotic systems by definition \citep{Devaney89}, it is useful to define the
\textit{deviation vector} as a point of the tangent bundle of the phase space
and interpret it as connecting two infinitesimally close trajectories. This 
vector evolves through the geodesic deviation equation 
\begin{equation}
\ddot{\xi}^\mu + \pder{{\Gamma^\mu}_{\kappa\lambda}}{x^\nu}\dot{x}^\kappa\dot{x}^\lambda\xi^\nu + 2{\Gamma^\mu}_{\kappa\lambda}\dot{x}^\kappa\dot{\xi}^\lambda = 0 \quad.
\end{equation}
As a measure of the deviation vector in a curved spacetime
\citep[see, e.g.,][]{Gerakopoulos14} we use 
\begin{equation}
\Xi^2 \defeq g_{\mu\nu}\xi^\mu\xi^\nu \quad.
\end{equation}
Typically, the deviation vector follows one of two behaviors - a linear one for
regular trajectories and an exponential one for chaotic trajectories. These
behaviors can be detected by the \textit{maximal Lyapunov characteristic exponent}
\begin{equation}
\mathrm{mLCE} \defeq \lim_{\tau\to\infty}\frac{1}{\tau}\log\left[\frac{\Xi\left(\tau\right)}{\Xi\left(0\right)}\right] \quad,
\end{equation}
which gives the inverse of a characteristic deviation time scale for chaotic
trajectories. In the case of regular trajectories, it behaves as
$\sim \tau^{-1}$ for large $\tau$, so in a plot in logarithmic scale it 
appears as a line of slope -1.

\begin{figure}
\begin{center}
\graphicspath{{}}
\input{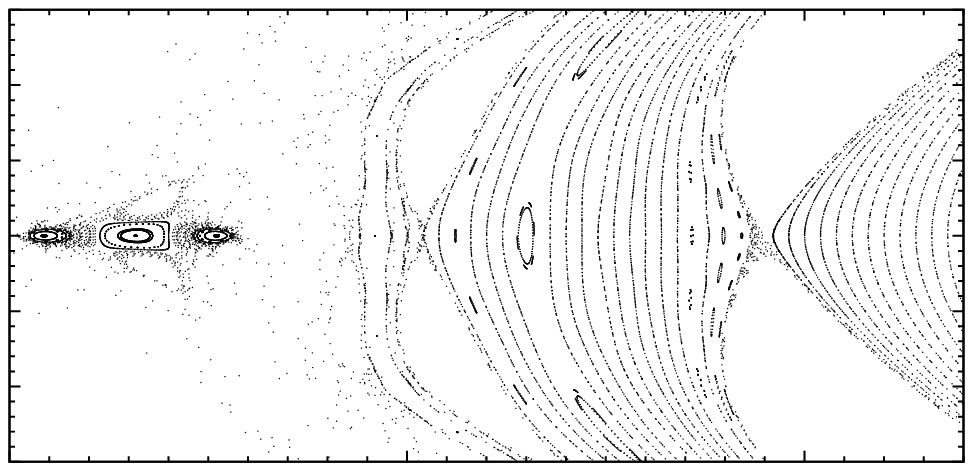}
\graphicspath{{}}
\input{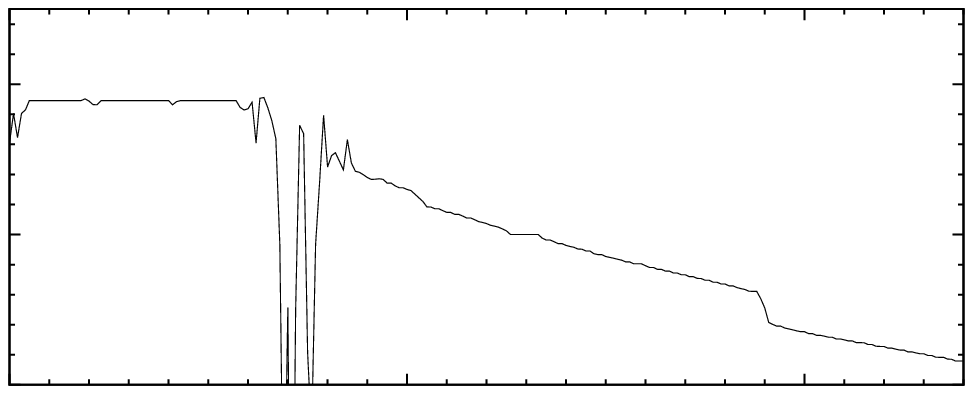}
\end{center}
\caption{\label{fig:surface_tip} 
Top panel: Detail of the left tip of the main island of stability.
Bottom panel: The rotation curve plotted for the $p_r=0$ line of the top panel.
The resonance plateaus along the curve are denoted by the respective fractions.}
\end{figure}

\begin{figure}
\begin{center}
\graphicspath{{}}
\input{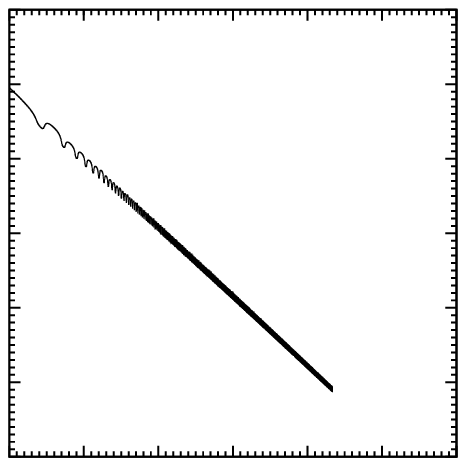}
\input{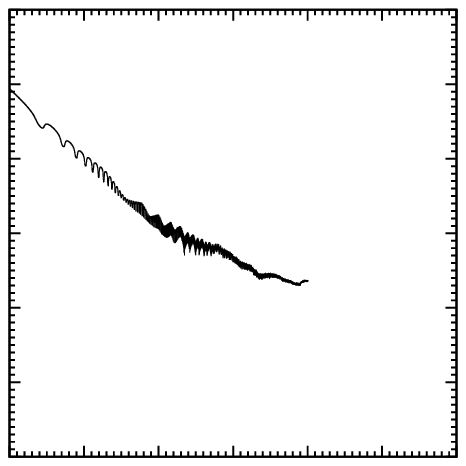}
\end{center}
\caption{\label{fig:deviation} Convergence of mLCE for a regular orbit (left
panel) with a chaotic orbit (right panel). Both orbits starting with $p_r=0$ on
the surface of section shown in Fig.~\ref{fig:surface_tip}, the regular from
$r=3.226$ and the chaotic from $r=3.2294$.}
\end{figure}

\section{Numerical examples}\label{sec:numerics}

All figures shown are plotted using the parameter values $M=m=1$, $a=0.5$,
$\epsilon_3 = 0.3$, $E = 0.95$, $L_z = 2.85$. In the left panel of 
Fig.~\ref{fig:projection_surface} are shown projections of two periodic orbits
on the meridian plane are shown, bounded by the CZV. In the right panel, the 
main island of stability in the Poincar\'{e} section is shown. The equatorial
plane $\theta = \pi/2$ with $\dot{\theta}>0$ is taken as the surface of section.
We notice no difference from an integrable system, as the chaotic behavior is
not prominent at this broad scale depiction. This difference becomes, however,
clearly visible in top panel of Fig.~\ref{fig:surface_tip}, which focuses on the
left tip of the main island of stability shown in the right panel of
Fig.~\ref{fig:projection_surface}. In particular, in the top panel of
Fig.~\ref{fig:surface_tip}, alongside with KAM curves, appear islands of
stability belonging to Birkhoff chains (ellipsoid-like structures) and chaotic
zones (scattered points). Under the panel containing this detail of the surface
of section, the corresponding rotation curve is plotted. The rotation curve
exhibits non-monotonic variations in a chaotic zone and plateaus (denoted by the
corresponding fraction) along islands of stability. Thus,
Fig.~\ref{fig:surface_tip} indicates that the JP spacetime corresponds to
a non-integrable system.

To directly estimate the sensitivity to initial conditions, we have calculated 
the mLCE. Fig. \ref{fig:deviation} shows the convergence of the mLCE for one
regular (left panel) and one chaotic orbit (right panel). For the regular orbit
indeed the mLCE convergence follows the -1 slope, while for the chaotic orbit  
the mLCE converges to a positive value.

\section{Conclusion}\label{sec:conclusion}

We have shown by numerical examples that geodesic motion in the JP spacetime
background corresponds to a non-integrable system, since chaos was detected.
The astrophysical implication is that if the spacetime around black holes is not
described by the Kerr metric, then one should expect imprints of chaos in
electromagnetic and gravitational wave signals coming from systems like extreme
mass ratio inspirals.

\ack

O.Z. and G.L-G are supported by Grant No. GACR-17-06962Y of the Czech Science
Foundation. We thank Petra Sukov\'{a} and Ond\v{r}ej Kop\'{a}\v{c}ek for useful
discussions.

\bibliography{\jobname}

\end{document}